\documentclass[iop]{emulateapj}

\def\psr{PSR~J0337}
\def\Tref#1{Table~\ref{tab:#1}}
\def\Fref#1{Figure~\ref{fig:#1}}
\def\Sref#1{Section~\ref{sec:#1}}

\newcommand{\kms}{\ensuremath{{\rm km}\,{\rm s}^{-1}}}
\newcommand{\logg}{\ensuremath{\log_{10}(g)}}
\newcommand{\Teff}{\ensuremath{T_{\rm eff}}}

\begin{document}

\title{Spectroscopy of the inner companion of the pulsar PSR J0337+1715\footnote{Based on observations obtained under Program GN-2012B-Q-43 at the
  Gemini Observatory, which is operated by the Association of
  Universities for Research in Astronomy, Inc., under a cooperative
  agreement with the NSF on behalf of the Gemini partnership: the
  National Science Foundation (United States), the National Research
  Council (Canada), CONICYT (Chile), the Australian Research Council
  (Australia), Minist\'{e}rio da Ci\^{e}ncia, Tecnologia e
  Inova\c{c}\~{a}o (Brazil) and Ministerio de Ciencia, Tecnolog\'{i}a
  e Innovaci\'{o}n Productiva (Argentina).  }} 
\slugcomment{ApJL, in press}

\author{David L.~Kaplan\altaffilmark{1}, 
Marten H.\ van~Kerkwijk\altaffilmark{2}, Detlev Koester\altaffilmark{3},
Ingrid H.\ Stairs\altaffilmark{4}, Scott M.~Ransom\altaffilmark{5},
Anne M.\ Archibald\altaffilmark{6,7}, Jason
W.~T.\ Hessels\altaffilmark{6,8}, Jason Boyles\altaffilmark{9,10}}
\altaffiltext{1}{Department of Physics, University of
Wisconsin-Milwaukee, 1900 E. Kenwood Boulevard, Milwaukee, WI 53211,
USA; kaplan@uwm.edu}
\altaffiltext{2}{Department of Astronomy and Astrophysics, University of Toronto, 50 St. George St., Toronto, ON, M5S 3H8, Canada; mhvk@astro.utoronto.ca}
\altaffiltext{3}{Institut f\"ur Theoretische Physik und Astrophysik, University of Kiel, 24098 Kiel, Germany}
\altaffiltext{4}{Department of Physics and Astronomy, University of
  British Columbia, 6224 Agricultural Road, Vancouver, British Columbia V6T 1Z1, Canada}
\altaffiltext{5}{National Radio Astronomy Observatory, 520 Edgemont Road, Charlottesville, Virginia 22903-2475, USA}
\altaffiltext{6}{ASTRON, the Netherlands Institute for Radio Astronomy, Postbus 2,
7990 AA, Dwingeloo, The Netherlands}
\altaffiltext{7}{Department of Physics, McGill University, 3600 rue University, Montreal, Quebec H3A 2T8, Canada}
\altaffiltext{8}{Astronomical Institute `Anton Pannekoek', University of Amsterdam,
Postbus 94249, 1090 GE Amsterdam, The Netherlands}
\altaffiltext{9}{Department of Physics and Astronomy, West Virginia University, White Hall, Box 6315, Morgantown, West Virginia 26506-6315, USA}
\altaffiltext{10}{Physics and Astronomy Department, Western Kentucky University, 1906 College Heights Boulevard \#11077, Bowling Green, Kentucky 42101-1077, USA}

\begin{abstract}
The hierarchical triple system PSR~J0337+1715 offers an unprecedented
laboratory to study secular evolution of interacting systems and to
explore the complicated mass-transfer history that forms millisecond
pulsars and helium-core white dwarfs.  The latter in
particular, however, requires knowledge of the properties of the individual
components of the system.  Here we present precise optical
spectroscopy of the inner companion in the PSR~J0337+1715 system.  We
confirm it as a hot, low-gravity DA white dwarf with
$\Teff=15,800\pm100\,$K and $\logg=5.82\pm0.05$. We also measure an
inner mass ratio of $0.1364\pm0.0015$, entirely consistent with that 
inferred from pulsar timing, and a systemic radial velocity of $29.7\pm0.3\,\kms$.  Combined with the
mass ($0.19751\,M_\odot$) determined from pulsar timing, our
measurement of the surface gravity implies a radius of
$0.091\pm0.005\,R_\odot$; combined further with the effective
temperature and extinction, the photometry implies a distance of
$1300\pm80\,$pc.  The high temperature of the companion is somewhat puzzling: with current models, it 
likely requires a recent period of unstable hydrogen burning, and
suggests a surprisingly short lifetime for objects at this phase in
their evolution.  We discuss the implications of these measurements in
the context of understanding the PSR~J0337+1715 system, as well as of low-mass
white dwarfs in general.
\end{abstract}

\section{Introduction}\label{sec:intro}

White dwarfs (WDs) are among the best-understood stars, enabling their
use as astrophysical tools in investigations of, e.g., the ages and
masses of astrophysical systems \citep[e.g.,][]{acigb10}.  However,
the lowest-mass WDs with He cores --- Extremely-Low Mass (ELM) WDs
with masses $<0.2\,M_\odot$ \citep[][and references
  therein]{kbap+12,bkap+13} --- still defy complete understanding,
with few reliable independent measurements of masses, sizes and ages
known \citep[e.g.,][]{vkbk96,bsp+05}.

Yet, these properties are  important for understanding the
evolution of ELM WDs and the binaries they are found in \citep*{il93,mdd95}.
For instance, while one would naively expect low-mass WDs to cool
quickly, given their relatively large size and small heat capacity,
some ELM WDs can remain bright and hot \citep[cf.][]{llfn95} because
they have outer hydrogen layers sufficiently thick for nuclear fusion
to continue -- stably or otherwise -- for several Gyr
(\citealt{asvdhp96,sarb02,pach07}; \citealt*{ambc13}).  Improving our understanding of
ELM WD cooling would aid in evolutionary models for, e.g., millisecond
pulsars and the later stages of mass transfer
\citep[e.g.,][]{tlk12,avkk+12,afw+13,kbvk+13}. Similarly, improved masses and
radii would aid in determining the final fates of double-WD binaries
(\citealt{db03}; \citealt*{mns04}; \citealt{davb+06}; \citealt*{kbs12}):
R~CrB stars, AM~CVn binaries, or even SNe Ia \citep{it84,webbink84}.

\object[PSR J0337+1715]{PSR~J0337+1715} (hereafter \psr;
\citealt{rsa+14}) was discovered in the 350\,MHz Green Bank Telescope
Driftscan survey \citep{blr+13,lbr+13}, and initial timing
observations found a $2.7\,$ms spin period, a $1.6\,$d orbital period,
and a likely companion mass of 0.1--0.2\,$M_\odot$, all of which are
consistent with expectations for a fully-recycled pulsar with a
low-mass He WD companion \citep{vkbjj05,tlk12}.  However, further
deviations to the observed spin period soon became apparent,
suggesting the presence of an additional body in the system.  This was
confirmed with an intensive timing campaign, finding an outer orbital
period of 327\,d \citep{rsa+14}.  By comparing the pulsar's pulse
arrival times with numerical integrations of possible orbits, the
masses, inclinations, and orbital parameters of the system could be
determined precisely \citep[][Archibald et al.\ 2014, in
  prep.]{rsa+14}, with a pulsar mass of $1.4378\pm0.0013\,M_\odot$, an
inner companion mass of $0.19751\pm0.0015\,M_\odot$, and an outer
companion mass of $0.4101\pm0.0003\,M_\odot$; both inner and outer
orbits are inclined by about $39\degr$ to the plane of the sky.

Before the nature of the system was fully determined, we identified an
unusually blue object coincident with it, which, based on initial
spectroscopy and photometry, we identified as the inner companion,
almost certainly a hot low-mass WD \citep{rsa+14}.  The brightness
of the system, its proximity, and the detailed constraints offered by
pulsar timing make the system a fantastic laboratory to explore the
atmosphere, structure, and evolution of ELM WDs.  Here, we present an
intensive spectroscopic campaign aimed to do so, which also serves as
a valuable cross-check of the  pulsar timing results.

\section{Observations}\label{sec:obs}

Spectra of the counterpart of \psr\ were taken for us between 2012
November and 2013 January with the Gemini Multi Object Spectrograph
(GMOS; \citealt{hook+04}) of the Gemini-North telescope (see
Table~\ref{tab:log}).  We used the $1200{\rm\,line\,mm^{-1}}$ grating,
covering the 3500--5000\,\AA\ range with the three $2048\times4608$
pixel CCD detectors (which were binned $4\times 4$, giving a spatial
scale of $0\farcs29{\rm\,pix^{-1}}$ and a dispersion of
$0.94\,{\rm\AA\,pix^{-1}}$).  As our object is relatively bright and
has broad absorption lines, we could use poor-seeing conditions and
hence opted for a wide, $1\farcs5$ slit.  With the typical seeing of
$1\farcs2$, and given the anamorphic plate scale of
$0\farcs38{\rm\,pix^{-1}}$ (for our grating setting;
\citealt{muro+03}), the resolution is $\sim\!3{\rm\,\AA}$.

In each visit, we took two 10-minute exposures offset by
$50{\rm\,\AA}$ to cover the gaps between the detectors.  Between the
exposures, we took incandescent and CuAr lamp spectra for both
settings, and before and after we took images through the slit to be
able to constrain velocity offsets due to centering errors.  For flux
calibration, spectra were taken on a photometric night using the same
settings through a $5\arcsec$ slit, immediately followed by exposures
of the white-dwarf spectrophotometric standard GD~71 \citep{bohlg04}.

We oriented the slit to include another
object on the slit, hoping to use it as a local flux and velocity
reference.  Unfortunately, the first object we picked -- the only
relatively bright one nearby, at a separation of $27\farcs7$ and position
angle $8\fdg3$ -- turned out to be a galaxy, which is not useful as it
fills the slit.  Hence, after the first set of data, we tried a
another object (at $58\arcsec$, $-112\degr$) which was 
fainter but which we hoped would have sufficiently narrow lines to still
be useful.  This, however, turned out to be a quasar.  Hence, as will
be clear below, our final velocity uncertainties have a significant component due to
slit centroiding errors.

We reduced the data using custom python scripts.  First, we subtracted
bias levels as determined from overscan regions for each of the six
read-out channels, divided by their gains to give electron counts
(with gains adjusted to ensure counts for flat fields were consistent
between different read-out channels), combined detector halves, and divided by
normalized flat fields.  The spectra were extracted optimally, fitting
at each dispersion position the trace of the two objects with a Moffat
function of the form $(1+[(x-x_c)/w]^2)^{-\beta}$ and the sky with a
second-degree polynomial (with the trace position $x_c$ and width $w$
allowed to vary slowly with wavelength, and the exponent fixed to
$\beta=2.5$).

For wavelength calibration, we first obtained accurate calibrations
for a set of daytime CuAr spectra taken through a narrow, $0\farcs5$
slit, in which many fewer lines are blended.  Fitting a third-order
polynomial for wavelength as a function of detector position,
optimizing simultaneously for the offsets between the chips, we find
root-mean-square residuals of $0.017{\rm\,\AA}$ (for 79 lines;
relative to copper and argon line wavelengths from the NIST database).
For each nighttime arc, we measure the shift relative to the daytime
arc taken at the same setting, and then apply the daytime calibration.

The spectra were flux-calibrated in three steps.  First, all spectra
were corrected approximately for atmospheric extinction using the
average Mauna Kea extinction curve from \citet{buto+13}.  Next, for the
narrow-slit spectra, slit and cloud losses were measured by fitting a
quadratic function to the count-rate ratio with the wide-slit
observation.  Finally, we divided by the instrumental response derived
from the smoothed ratio of the count rates and fluxes of GD~71.  While
the wavelength region we covered does not fully overlap any of the
photometric filters, extrapolating the
short-wavelength part of the spectrum using a power-law gives a good
agreement (better than 1-$\sigma$) with the measured SDSS $u^\prime$ photometry.

\section{Analysis}

\subsection{Velocities}

\begin{figure*}
\plotone{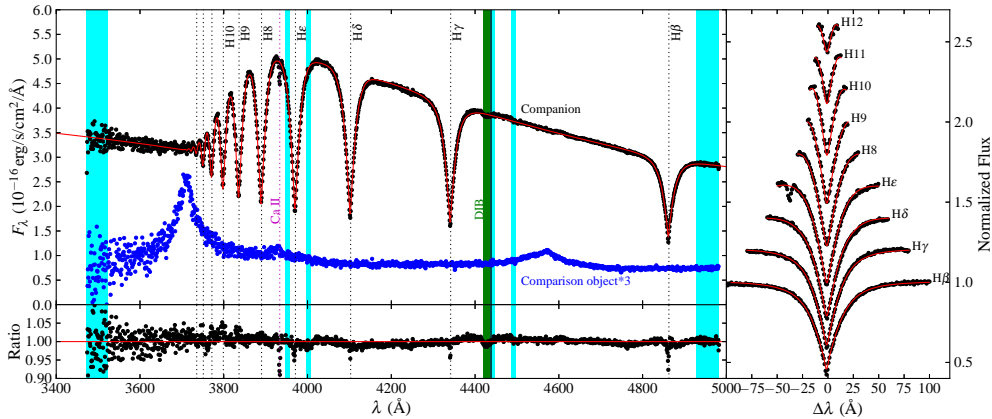}
\caption{Left: Composite spectrum of the optical counterpart to \psr.  The
 individual spectra have been shifted by their measured velocities
 and summed.  Overdrawn is the best-fit model atmosphere (red),
 with effective temperature 15,800\,K and surface gravity
 $\log(g)=5.82$.  We also show the comparison source below the
 counterpart (multiplied by 3 for clarity), which we  identified
 as a $z=1.4$ quasar: the broad emission line at 3717\,\AA\ is
 \ion{C}{4} $\lambda$1549.  The vertical lines mark the Balmer
 series, with \ion{Ca}{2} $\lambda$3933 also labeled.  The cyan bands
 show where we have data from only half of the spectra, either
 because of the chip gaps or the ends of the spectral coverage.  The
 green band shows the diffuse interstellar band (DIB) at 4430\,\AA.
 Right: fits to the individual Balmer lines (as labeled), with the
 model overdrawn in red.
 Bottom:  the ratio of the counterpart spectrum with respect
 to the model: small deviations are seen at some of the lower-order
 Balmer lines, with a more significant deviation at the \ion{Ca}{2}
 $\lambda$3933 line.}
\label{fig:composite}
\end{figure*}

We determined velocities by fitting a template to the data for a range
of trial velocities, at each allowing for normalization and possible
variations with wavelength using a linear function.  For the template,
we used a pure hydrogen model atmosphere with $\Teff=15,800$\,K and
$\logg=5.80$, close to the best-fit parameters determined below
(\Sref{model}), convolving it with a Gaussian with a width set by the
typical seeing of $1\farcs2$ (equivalent to $2.9{\rm\,\AA}$), and
truncated at $1\farcs5$ to mimic the slit.  The fits to the spectra
were good, with typical $\chi^2=1440$ for 1527 degrees of freedom, and
implied formal velocity uncertainties of $\sim\!4{\rm\,km\,s^{-1}}$.

An additional uncertainty in our velocities is the extent to which the
object was properly centered in the slit.  As mentioned, we had hoped
to use a comparison star to calibrate this, but this turned out to be
a quasar.  Inspecting the images taken through the slit before and
after the spectra, we find that the star has root-mean-square offsets
from the slit center of $0\farcs07$, with the largest deviations about
twice that (i.e., up to 2 unbinned pixels).  The effect on velocity
also depends on seeing (for very bad seeing, the slit is better filled
and the effect minimized).  From the acquisition images themselves, we
find flux-weighted offsets about half as large, of $0\farcs045$ (rms).
The shifts for the spectra are slightly smaller, since the traces in
the spectra show slightly larger seeing, presumably because of
jitter in the telescope pointing.  Using the centroiding positions
from the acquisition images combined with the seeing from the spectral
traces, we infer flux-weighted offsets of $0\farcs034$ (rms),
corresponding to wavelength shifts of $0.060{\rm\,\AA}$ and, for an
assumed effective wavelength of 4200\,\AA, velocity shifts of
$\sim\!4.3{\rm\,km\,s^{-1}}$.  In \Tref{log}, we list the velocities
corrected for these shifts and corrected to the Solar System
barycenter.

To determine the orbit, we fit the velocities to a model $v_{\rm
  WD}(t) = \gamma - (1/q_I)\times v^{\rm in}_{\rm PSR}(t) + v^{\rm
  out}_{\rm PSR}(t)$, where $\gamma$ is the systemic radial
velocity, $q_I\equiv M_{\rm WD}/M_{\rm PSR}$ is the inner mass ratio,
 $v^{\rm in}_{\rm PSR}$ is the radial
velocity of the pulsar in the inner orbit, and  $v^{\rm out}_{\rm
  PSR}$ is the radial velocity of the inner-orbit barycenter in the
outer orbit, both as inferred from
the timing model (with radial-velocity amplitudes $K^{\rm in}_{\rm
  PSR}=16.291$ and $K^{\rm out}_{\rm PSR}=4.978{\rm\,km\,s^{-1}}$,
respectively).  Such a model is traditional for pulsar binaries, where
the period and pulsar's velocity are known but the mass ratio is not.
For \psr, we do know $q_I$, but the analysis serves as a valuable
check of the full timing model.  The velocities fit reasonably well
(\Fref{orbit}), although the fit is formally unacceptable, with
$\chi^2=55.9$ for 33 degrees of freedom (35 spectra, 2 parameters),
likely because of remaining uncertainties in the slit centering (e.g.,
due to differential atmospheric refraction).  In order to account for
this in our parameter uncertainties, we added an additional
uncertainty of $3.3{\rm\,km\,s^{-1}}$ to the velocity errors (in
quadrature), giving a reduced $\chi^2$ of 1.0.  With that, we find
$\gamma=29.7\pm0.9\,{\rm km\,s}^{-1}$ and $q_I=0.1364\pm0.0015$,
implying $K^{\rm in}_{\rm WD}=119.4\pm1.3{\rm\,km\,s^{-1}}$.  This is
fully consistent with the value measured from pulsar timing
\citep{rsa+14}: $q_I=0.13737\pm0.00004$.

For completeness, we note that our results do not depend on whether we
include the centroiding shifts, although the fit becomes substantially
worse if we do not ($\chi^2=115$).  The fit also does not depend on
whether we took the centroid from the nearest acquisition image, or
rather interpolated between images taken before and after the spectra.
Furthermore, if we leave the prefactor for the centroiding shift free,
i.e., include an additional term $\alpha\Delta v$ in the fit, we find
$\alpha=1.13\pm0.16$.  Finally, if we ignore the contribution from the
outer orbit, we find $\chi^2=58.5$, i.e., the outer orbit is detected
marginally even in our velocities.

\begin{figure}
\plotone{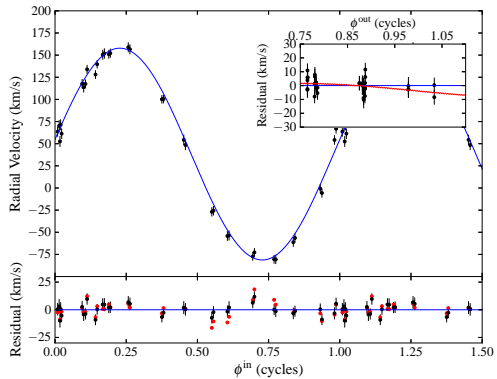}
\caption{ Radial velocities of the optical counterpart to \psr.  {\em
   Main panel:} Observed velocities plus the velocity expected for
 the outer orbit, as a function of the phase of the inner orbit
 (repeated 1.5 times for clarity), with the best-fit model overdrawn
 (solid line).  {\em Lower panel:} Residuals with respect to that
 model with and without correction for small slit-centering errors
 (black and red points, respectively).  {\em Inset:} Residuals as a
 function of outer orbital phase, with inner velocity contribution
 removed, and with models with and without the outer object overdrawn
 (blue and red solid lines, respectively).  }
\label{fig:orbit}
\end{figure}

\subsection{Model atmosphere fits}
\label{sec:model}
Given the velocities determined above, we created a composite summed
spectrum (\Fref{composite}) by shifting each of the individual
measurements back to zero velocity.  We see strong Balmer lines, as
well as a weaker \ion{Ca}{2} $\lambda$3933 absorption line and some
broad absorption near 4430\,\AA\ associated with a diffuse
interstellar band (DIB).

Metal lines are occasionally seen from white dwarfs \citep{gkf+12},
especially those with low gravities ($<5.6$;
\citealt{kbap+12,kbvk+13}), and they are commonly interpreted as signs
of accretion.  However, \ion{Ca}{2} absorption can also be
interstellar in origin.  We compared the velocity centroid and width
of the \ion{Ca}{2} line using both the spectra corrected for the
motion of the white dwarf and only corrected for the motion of the
Earth around the Solar System barycenter.  We find that the line is
marginally narrower when not correcting for the motion of the white
dwarf (the depth increases from $9.3\pm0.6$\% to $10.2\pm0.6$\%),
suggesting that it is interstellar in this case.  This is confirmed by
a line of similar strength that we see in the spectrum of the
comparison quasar, although the signal-to-noise is lower.  The
centroid of the \ion{Ca}{2} line as determined from the uncorrected
data is $44\pm6\,\kms$, implying a systemic radial velocity of the
pulsar relative to the interstellar medium of $-14\pm6\,\kms$; this is
not very different from the expectations using the \citet{bb93}
rotation curve for a distance of 1.3\,kpc (see below), where the
radial velocity goes between 0 and $-4\,\kms$ along the line-of-sight.
Similarly, with data only corrected for the Earth's motion (and
excluding data where the gap
between the green and red CCDs came near the DIBs), we measure
a depth at the center of the DIB of $2.0\pm0.2$\%.  Based on the
empirical relation of \citet{krelwgh87}, we infer an extinction
$A_V\approx 0.3\,$mag.

We next fit for the atmospheric parameters for the white dwarf by
comparing against pure hydrogen models computed by one of us
(D.~Koester).  These 
models covered $\Teff=15,500\,$K to 16,500\,K in steps of 100\,K and
$\logg=5.50$ to 6.50 in steps of 0.1\,dex.  We identified some
line-free regions to fit a cubic polynomial that represented the
difference in normalization between the models and the data, and
computed the $\chi^2$ of each model with respect to the data.  The
models were convolved with a function to represent the slit and the
average seeing ($1\farcs$2), as discussed above.  We excluded the
region around the \ion{Ca}{2} $\lambda$3933 line and H$\epsilon$
(which is blended with \ion{Ca}{2} $\lambda$3968).  Overall, our
initial 
fit has $\Teff=15,780\,$K and $\logg=5.82$.  This fit has
$\chi^2=4098.2$ for 1613 degrees-of-freedom, so it is formally
unacceptable.  Much of the deviation comes from the cores of the lower-order
Balmer lines (\Fref{composite}, lower panel).  These deviations might
be indicative of the outer member of the binary (i.e., third light).
However, we find that they track the white dwarf's orbit, so are
likely just errors in our model or calibration.  In particular, NLTE
effects might be important; in higher-gravity white dwarfs, they cause
deeper cores for H$\alpha$ and H$\beta$ at these temperatures.

Computing the best-fit model for each individual observation gave
similar results, with $\Teff=15,882\pm35\,$K and $\logg=5.85\pm0.01$,
where the uncertainties are the formal errors in the means.  To account
for the formally poor fit and model uncertainties, we increase the
uncertainties to $\pm0.05$\,dex and $\pm100\,$K, which are about the
smallest we would believe for an object in this relatively
unconstrained part of the white dwarf cooling sequence.  We therefore
adopt as our best-fit model $\Teff=15,800\pm100\,$K and
$\logg=5.82\pm0.05$.  For this effective temperature, the best-fit
extinction based on the photometry is $A_V=0.44\pm0.04\,$mag \citep{rsa+14}.

\section{Discussion \& Conclusions}
Our measurements provide the velocities and atmospheric parameters of
the inner white dwarf in the \psr\ system.  The velocities serve to confirm the more
precise ones inferred from pulsar timing.  In addition, they show that
the systemic velocity is low.  This is not unexpected, since any kick
imparted to the system in the supernova explosion that formed the
neutron star must have been small for the triple to survive 
\citep{tvdh14}.  One
thus expects the proper motion to be similarly small.

Our measurement of the surface gravity, combined with the precise mass
from timing, implies a radius of $0.091\pm0.005\,R_\odot$.  Combining
this in turn with the effective temperature, extinction, and
photometry, one infers a distance of $1300\pm80\,$pc \citep{rsa+14}.
With an accurate parallax from very-long baseline interferometry
(measurements are in progress), this can be used to infer the surface
gravity and thus test the model atmospheres in an otherwise poorly
constrained regime.

The mass and radius of the inner white dwarf are consistent with those
expected for a young, low-mass helium-core white dwarf, similar to the
white dwarf companions found in other binaries
(Section~\ref{sec:intro}).  Compared to low-mass white dwarfs around
pulsars,\footnote{One cannot easily compare  low-mass white dwarfs
 with white-dwarf or A-star companions, since for those systems there
 are strong observational biases to find hotter, more luminous and
 larger white dwarfs.} however, the source stands out for being
hotter than most.  This must be intrinsic as possible contributions
from pulsar irradiation and tidal heating are negligible.

The high temperature is surprising as it would suggest the system is
in a short-lived state and hence that similar systems are common --
which, empirically, they are not.  This suggestion arises because in
current evolutionary models of helium white dwarfs, temperatures in
excess of $\sim\!12,000\,$K are only achieved in models with unstable
shell flashes (e.g., \citealt{dsbh98}; \citealt*{asb01}).  In those
shell flashes, however, most of the thick hydrogen layer is lost, and
hence the white dwarf will cool relatively quickly.  Furthermore,
while the flashing state may last $\sim\!200\,$Myr \citep{ambc13}, the
typical cooling timescales at these temperatures are short, a few
$10\,$Myr for each flash, and hence the total time spent at high
temperatures is often $<100\,$Myr, depending on mass.  These
timescales are still far longer than the expected sedimentation
timescale for helium ($\sim 10^3\,$yr) following mixing during a shell
flash, consistent with the lack of any \ion{He}{1} in the spectrum of the inner
WD (cf.\ \citealt{kbvk+13}).  Based on inspection, we can
roughly limit its abundance to $10^{-2.5}$H (by number), which would
not change our inferred
\logg\ by more than our quoted uncertainty.  If the inner white dwarf
is really only a few 100\,Myr old, it almost certainly formed last, as
also expected from simple models \citep{rsa+14,tvdh14}, although we
cannot strictly exclude the opposite: at the upper limit to the
temperature of the outer WD of 20,000\,K \citep{rsa+14}, the cooling
time would be 30--100\,Myr.

Of course, it could be a coincidence that we found such a hot
white-dwarf companion.  However, a similarly hot companion was found
for \object[PSR J1816+4510]{PSR~J1816+4510} \citep{kbvk+13}.  Since
typical millisecond pulsars remain visible for a Hubble time, and
since we know $\sim\!50$ pulsars with low-mass white dwarf companions,
this suggests that white dwarfs can stay hot for $\sim\!500\,$Myr,
substantially longer than expected in current theoretical models.  The
discrepancy is made worse by the fact that white dwarfs with mass
below $\lesssim\!0.18\,M_\odot$ are not expected to get this hot at
all (they should not have flashes).  Observationally, however,
lifetimes of a few hundred Myr seem also indicated by the prevalence
of hot, low-mass white dwarfs around A stars (which have ages of
$\lesssim\!1\,$Gyr).

\acknowledgements{We thank an anonymous referee for a helpful
  suggestion. We thank Gemini staff Atsuko Nitta, John Blakeslee, and
  Kristin Chiboucas for helping assure the high quailty of these data.
  The National Radio
  Astronomy Observatory is a facility of the National Science
  Foundation, operated under cooperative agreement by Associated
  Universities, Inc.  Balmer/Lyman lines in the
  models were calculated with the modified Stark broadening profiles
  of \citet{tb09}, kindly made available by the authors.
  M.H.v.K.\ and I.H.S.\ acknowledge funding from NSERC Discovery
  Grants.  J.W.T.H.\ and A.M.A.\ acknowledge support from a Vrije
  Competitie grant from NWO.  We made extensive use of
SIMBAD, ADS, and  Astropy
  (\url{http://www.astropy.org}; \citealt{astropy}).}

{\it Facilities:} \facility{Gemini:Gillett (GMOS)}



\begin{deluxetable*}{lcccccccccc}
\tablewidth{0pt}
\tablecaption{Log of observations and velocity measurements\label{tab:log}}
\tablehead{\colhead{Date} & \colhead{UT} & \colhead{$\lambda_c\tablenotemark{a}$} & \colhead{Seeing} & \colhead{Shift\tablenotemark{b}} & \colhead{Offset\tablenotemark{c}} & \colhead{${\rm MJD_{bar}}$} & \colhead{$\phi^{\rm in}$\tablenotemark{,d}} & \colhead{$v^{\rm in}_{\rm PSR}$\tablenotemark{e}} & \colhead{$v^{\rm out}_{\rm PSR}$\tablenotemark{f}} & \colhead{$v_{\rm WD}$\tablenotemark{g}}\\ \colhead{None} & \colhead{None} & \colhead{(\AA)} & \colhead{(arcsec)} & \colhead{(arcsec)} & \colhead{(arcsec)} & \colhead{None} & \colhead{None} & \colhead{$({\rm km\,s^{-1}})$} & \colhead{$({\rm km\,s^{-1}})$} & \colhead{$({\rm km\,s^{-1}})$}}
\startdata
2012 Nov 09\ldots & 08:44 & 4250 & $1.4$ & $-0.024$ & $-0.010$ & 56240.3674 & 0.0918 & $-10.68$ & \phs$4.92$ & \phs$117\pm4$ \\
& 08:56 & 4200 & $1.3$ & $-0.055$ & $-0.025$ & 56240.3760 & 0.0971 & $-11.09$ & \phs$4.92$ & \phs$112\pm5$ \\
& 09:10 & 4250 & $1.3$ & $-0.055$ & $-0.025$ & 56240.3863 & 0.1034 & $-11.55$ & \phs$4.92$ & \phs$117\pm4$ \\
& 09:24 & 4200 & $1.2$ & $-0.057$ & $-0.029$ & 56240.3956 & 0.1091 & $-11.96$ & \phs$4.92$ & \phs$133\pm4$ \\
& 11:32 & 4250 & $1.9$ & \phs$0.034$ & \phs$0.009$ & 56240.4850 & 0.1640 & $-15.00$ & \phs$4.92$ & \phs$149\pm4$ \\
& 11:46 & 4200 & $1.9$ & \phs$0.036$ & \phs$0.009$ & 56240.4945 & 0.1698 & $-15.22$ & \phs$4.92$ & \phs$149\pm5$ \\
2012 Nov 14\ldots & 07:52 & 4250 & $1.0$ & $-0.047$ & $-0.028$ & 56245.3318 & 0.1387 & $-13.79$ & \phs$4.84$ & \phs$128\pm4$ \\
& 08:06 & 4200 & $0.9$ & $-0.047$ & $-0.030$ & 56245.3413 & 0.1445 & $-14.10$ & \phs$4.84$ & \phs$139\pm4$ \\
& 12:22 & 4250 & $1.8$ & $-0.021$ & $-0.006$ & 56245.5193 & 0.2537 & $-16.08$ & \phs$4.84$ & \phs$159\pm5$ \\
& 12:36 & 4200 & $1.8$ & \phs$0.017$ & \phs$0.005$ & 56245.5288 & 0.2596 & $-15.98$ & \phs$4.84$ & \phs$155\pm5$ \\
2012 Nov 15\ldots & 08:23 & 4250 & $1.2$ & $-0.133$ & $-0.067$ & 56246.3536 & 0.7658 & \phs$15.85$ & \phs$4.82$ & \phn$-81\pm4$ \\
& 08:37 & 4200 & $1.5$ & $-0.131$ & $-0.050$ & 56246.3632 & 0.7717 & \phs$15.70$ & \phs$4.82$ & \phn$-81\pm4$ \\
& 11:02 & 4250 & $1.6$ & $-0.060$ & $-0.021$ & 56246.4636 & 0.8333 & \phs$12.89$ & \phs$4.82$ & \phn$-61\pm4$ \\
& 11:15 & 4200 & $1.5$ & $-0.050$ & $-0.020$ & 56246.4731 & 0.8392 & \phs$12.51$ & \phs$4.82$ & \phn$-57\pm4$ \\
2012 Nov 16\ldots & 08:05 & 4250 & $0.8$ & $-0.045$ & $-0.034$ & 56247.3409 & 0.3718 & $-10.11$ & \phs$4.79$ & \phs$100\pm4$ \\
& 08:19 & 4200 & $0.8$ & $-0.052$ & $-0.039$ & 56247.3505 & 0.3776 & \phn$-9.63$ & \phs$4.79$ & \phs$100\pm4$ \\
2012 Dec 15\ldots & 08:45 & 4250 & $1.3$ & $-0.065$ & $-0.030$ & 56276.3683 & 0.1858 & $-15.72$ & \phs$3.33$ & \phs$150\pm4$ \\
& 08:58 & 4200 & $1.3$ & $-0.027$ & $-0.013$ & 56276.3772 & 0.1912 & $-15.86$ & \phs$3.32$ & \phs$151\pm4$ \\
2012 Dec 17\ldots & 10:09 & 4250 & $2.2$ & $-0.051$ & $-0.011$ & 56278.4272 & 0.4491 & \phn$-2.93$ & \phs$3.17$ & \phs\phn$ 54\pm5$ \\
& 10:23 & 4200 & $2.1$ & $-0.078$ & $-0.018$ & 56278.4367 & 0.4550 & \phn$-2.34$ & \phs$3.17$ & \phs\phn$ 47\pm5$ \\
2012 Dec 18\ldots & 04:54 & 4250 & $1.2$ & \phs$0.011$ & \phs$0.005$ & 56279.2084 & 0.9285 & \phn\phs$4.98$ & \phs$3.11$ & \phn\phn$ -1\pm4$ \\
& 05:08 & 4200 & $1.2$ & $-0.013$ & $-0.006$ & 56279.2180 & 0.9343 & \phn\phs$4.41$ & \phs$3.11$ & \phn\phn$ -6\pm4$ \\
& 08:16 & 4250 & $2.4$ & $-0.115$ & $-0.021$ & 56279.3484 & 0.0144 & \phn$-3.71$ & \phs$3.10$ & \phs\phn$ 52\pm5$ \\
& 08:29 & 4200 & $2.2$ & $-0.177$ & $-0.036$ & 56279.3580 & 0.0202 & \phn$-4.29$ & \phs$3.10$ & \phs\phn$ 60\pm5$ \\
2012 Dec 19\ldots & 05:07 & 4250 & $1.2$ & \phs$0.075$ & \phs$0.036$ & 56280.2176 & 0.5477 & \phn\phs$6.92$ & \phs$3.03$ & \phn$-27\pm4$ \\
& 05:21 & 4200 & $1.3$ & \phs$0.070$ & \phs$0.032$ & 56280.2272 & 0.5535 & \phn\phs$7.46$ & \phs$3.03$ & \phn$-26\pm4$ \\
& 07:16 & 4250 & $1.6$ & \phs$0.114$ & \phs$0.041$ & 56280.3068 & 0.6024 & \phs$11.48$ & \phs$3.02$ & \phn$-55\pm4$ \\
& 07:30 & 4200 & $1.9$ & \phs$0.119$ & \phs$0.033$ & 56280.3164 & 0.6083 & \phs$11.90$ & \phs$3.02$ & \phn$-55\pm5$ \\
& 10:45 & 4250 & $1.1$ & $-0.054$ & $-0.030$ & 56280.4519 & 0.6914 & \phs$15.86$ & \phs$3.01$ & \phn$-78\pm4$ \\
& 10:58 & 4200 & $1.0$ & $-0.089$ & $-0.054$ & 56280.4614 & 0.6973 & \phs$15.99$ & \phs$3.01$ & \phn$-74\pm3$ \\
2013 Jan 18\ldots & 07:33 & 4250 & $1.3$ & \phs$0.004$ & \phs$0.002$ & 56310.3181 & 0.0074 & \phn$-4.34$ & \phs$0.19$ & \phs\phn$ 63\pm5$ \\
& 07:46 & 4200 & $1.5$ & $-0.032$ & $-0.012$ & 56310.3272 & 0.0130 & \phn$-4.88$ & \phs$0.19$ & \phs\phn$ 69\pm6$ \\
& 07:58 & 4200 & $1.5$ & $-0.032$ & $-0.012$ & 56310.3357 & 0.0182 & \phn$-5.38$ & \phs$0.19$ & \phs\phn$ 70\pm5$ \\
2013 Feb 05\ldots & 05:33 & 4250 & $1.6$ & $-0.025$ & $-0.008$ & 56328.2349 & 0.9810 & \phn$-3.92$ & $-1.61$ & \phs\phn$ 55\pm4$ \\
& 05:46 & 4200 & $1.6$ & $-0.030$ & $-0.010$ & 56328.2438 & 0.9864 & \phn$-4.45$ & $-1.62$ & \phs\phn$ 67\pm4$ \\
\enddata

\tablenotetext{a}{Center wavelength of each observation.}
\tablenotetext{b}{Deviation of the target from the center of the slit
 in the acquisition image.}
\tablenotetext{c}{Deviation of the center of light from the center of
 the slit, calculated using the FWHM measured from the spectra.}
\tablenotetext{d}{Orbital phase for the inner orbit.}
\tablenotetext{e}{Inferred velocity for the pulsar due to the inner
 orbit relative to the system barycenter.}
\tablenotetext{f}{Inferred velocity for the pulsar due to the outer
 orbit relative to the system barycenter}
\tablenotetext{g}{Measured velocity for the white dwarf, with all
 corrections applied.}
\end{deluxetable*}

\end{document}